\documentclass[twocolumn,preprintnumbers,amsmath,amssymb]{revtex4}
\usepackage{graphicx}
\usepackage{dcolumn}
\usepackage{bm}
\def\be{\begin{equation}}
\def\ee{\end{equation}}
\def\bea{\begin{eqnarray}}
\def\eea{\end{eqnarray}}

\begin{document}
\preprint{CERN-TH-2019-152}


\title{A Standard Model explanation for the excess of electron-like events in MiniBooNE}

\author{A. Ioannisian}

\affiliation{ 
CERN, Theory Division, CH-1211 Geneva 23, Switzerland\\
Yerevan Physics Institute, Alikhanian Brothers 2, Yerevan-36, Armenia\\
 Institute for Theoretical Physics and Modeling, Yerevan-36, Armenia }

\begin{abstract}
We study the dependence of neutral current (NC) neutrino-induced $\pi^0$/photon production ($\nu_\mu + A \to \nu_\mu +1\pi^0 / \gamma + X$) on the atomic number of the target nucleus, A,  at 4-momentum transfers  relevant to the MiniBooNE experiment: $\Delta$ resonance mass region. 
Our conclusion is based  on experimental data for photon-nucleus interactions from the A2 collaboration at the Mainz MAMI accelerator. We work in the approximation that decays of $\Delta$ resonance unaffected by its production channel, via photon or Z boson. $1\pi^0+X$ production scales as A$^{2/3}$, the surface area of the nucleus. Meanwhile the photons 
created in $\Delta$ decays will leave the nucleus, and that cross section will be proportional to the atomic number of the nucleus. Thus the ratio of photon production to $\pi^0$ production is proportional to A$^{1/3}$. For carbon $^{12}$C this factor is $\approx$2.3. MiniBooNE normalises the rate of photon production to the measured $\pi^0$ production rate. The reduced neutral pion production rate would yield at least twice as many photons as previously expected, thus significantly lowering the number of unexplained electron-like events. 
\\
\end{abstract}
\maketitle
The MiniBooNE experiment searches  for $\stackrel{ { (-)}}{\nu_\mu} \to \stackrel{(-)}{\nu_e} $ oscillations. It reported a significant excess of low-energy (200-500 MeV) electron-like events in both neutrino \cite{Aguilar-Arevalo:2018gpe} and antineutrino\cite{Aguilar-Arevalo:2013pmq} runs. This anomaly gave rise to speculations of new physics beyond the Standard Model. 

It is well known that at these energies MiniBooNE cannot distinguish electrons/positrons from photons. The energy range 200-500MeV is the $\Delta$ resonance region and the dominant majority of photons will be produced in $\Delta$ decays.

The A2 collaboration \cite{A2} at the Mainz MAMI accelerator investigates photon absorption on nucleus/nucleons. The incident photon beam has a very well known energy, flux and polarisation, in energy range 40-1603 MeV, yielding  precise measurements of photon absorption cross sections on nucleus/nucleon with different final states.

MiniBooNE is a liquid-scintillator detector and uses methylene (H$_2$C=CH-CH$_3$). On average it is CH$_2$: 2 protons per carbon nucleus.

MiniBooNE estimates the possible number of produced photons via NC, $\nu_\mu +CH_2 \to \nu_\mu+\gamma +X$, by the number of detected $\pi^0$s from $\nu_\mu +CH_2 \to \nu_\mu+\pi^0 +X$. They assume that the ratio of produced photons to $\pi^0$s is almost a constant, largely independent of whether the photon/$\pi^0$ produced on carbon or on a free proton.

According to \cite{Tanabashi:2018oca} \footnote {That decay rate was not measured instead it is calculated from the measured amplitude of the $\gamma +p \to \Delta^+ $\cite{Tanabashi:2018oca}.}
\be
Br(\Delta^{+/0} \to p/n +\gamma)=(5.5 - 6.5) 10^{-3}
\ee
and the branching rate to produce $\pi^0$ is
\be
Br(\Delta^{+/0} \to p/n +\pi^0)\simeq 2/3.
\ee
Thus the probability to have produced a photon per produced $\pi^0$ is
\be
{\Gamma^\gamma (\Delta^{+/0})\over \Gamma^{\pi^0}(\Delta^{+/0})} \simeq (8.25 - 9.75) 10^{-3}.
\label{rr}
\ee
MiniBoone used this number to estimate the number of photons produced.


\begin{figure}[t!]
 \includegraphics[width=0.5\textwidth]{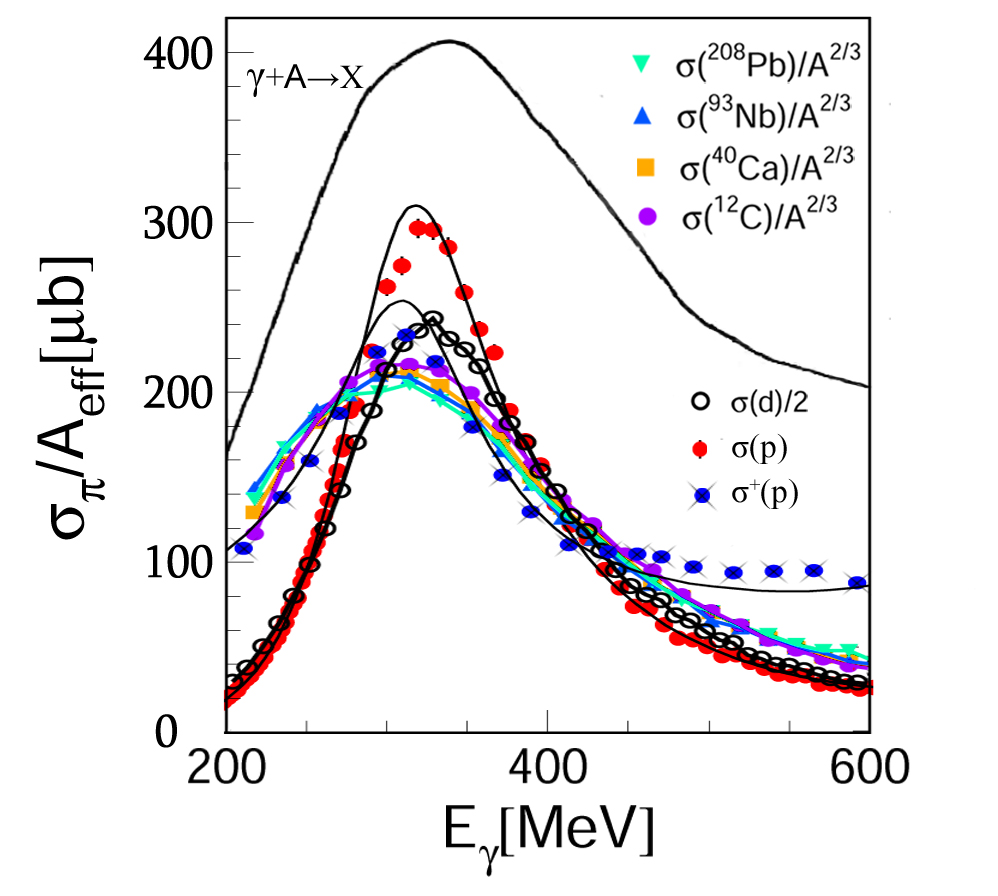}
\caption[...]{Inclusive $\pi^0$ production cross sections $\gamma +A \to \pi^0+X$. $A_{eff}$=1 for proton, 2 for deuteron  and $A^{2/3}$ for nuclei.  $\sigma^+$ is the cross section of $\gamma +p \to n + \pi^+$.  $\gamma + A\to X$  is (per nucleon) photo-absorption cross section on nuclei, $A_{eff}$=A. \cite{Krusche:2004xz}
\label{Fig0}}
\end{figure}

According to the A2 collaboration  \cite{Krusche:2004xz} the $\pi^0$ photo-production on nuclei scales as A$^{2/3}$
\be
\sigma(\gamma + A \to 1\pi^0 +X ) \propto A^{2/3}.
\ee
As it is well known the pion-nucleus interaction cross section is proportional to the surface area of the nucleus. By contrast the total photon absorption cross section on the nucleus is proportional to its volume and scales as A \cite{Krusche:2004xz}.
Meanwhile the photons created in $\Delta$ decays will leave the nucleus, and that cross section will be proportional to the atomic number of the nucleus.

Thus we conclude that the ratio of photon to $\pi^0$ production via a $\Delta$ resonance in nuclei is proportional to A$^{1/3}$.

Of course, this scaling is strictly correct only for nuclei, but for rough estimate we assume the same relation for protons as well \footnote {It is in our plan to make more accurate calculations by taking into account spectre of neutrinos as well.}.

For MiniBooNE via NC on carbon there will be about 2.3 times more photons than the naive estimation of eq \ref{rr}. Thus from CH$_2$ there will be at least twice more photons, reducing the significance of the excess of electron-like evens to just  2.2 $\sigma$.

Similarly at the short baseline ICARUS experiment at FERMILAB, one may expect that there will be at least 40$^{1/3} \simeq$3.4   times more photons per $\pi^0$ than one may expect from eq \ref{rr}.


We would like to thank Bernd Krusche and Georg Raffelt for discussions.



\end{document}